\documentclass[a4paper,11pt]{article}
\pdfoutput=1 

\usepackage{jheppub}
                     
\usepackage[T1]{fontenc}  
\graphicspath{{Figures/}} 
\usepackage{booktabs}     
\usepackage{bbold}        
\usepackage{mathtools}
\usepackage{multirow}
\usepackage{physics}
\usepackage[symbol]{footmisc}
\usepackage{cancel}
\usepackage{dsfont}
\usepackage{rotating}
\usepackage{amsmath}
\usepackage[outline]{contour}

\graphicspath{{Figures/}} 	     

\newcommand{\tstring}[2]{\texorpdfstring{#1}{#2}}
\DeclareMathOperator{\eV}{\text{eV}}
\DeclareMathOperator{\MeV}{\text{MeV}}
\DeclareMathOperator{\GeV}{\text{GeV}}
\DeclareMathOperator{\TeV}{\text{TeV}}

\DeclareMathOperator{\Z2}{\mathbb{Z}_2}

\title{\bf Neutrino Masses in the Mirror Twin Higgs with Spontaneous \tstring{$\mathbb{Z}_2$}{Z2} Breaking}

\author[a]{Pedro Bittar,} 
\author[b]{Davi B. Costa,}
\author[a]{Gustavo Burdman}
\affiliation[a]{Department of Mathematical Physics,
Institute of Physics, 
University of São Paulo, \\
R. do Matão 1371, São Paulo, 
SP 05508-090, Brazil \vspace{0.2cm}}

\affiliation[b]{Department of Physics and Enrico Fermi Institute, University of Chicago, \\
933 East 56th Street, Chicago, Illinois 60637, U.S.A.}

\emailAdd{bittar@if.usp.br}
\emailAdd{davicosta@uchicago.edu}
\emailAdd{gaburdman@usp.br}

\abstract{We introduce a mirror twin Higgs model with spontaneous $\mathbb{Z}_2$ symmetry breaking that ameliorates the constraints in twin Higgs cosmology and, at the same time, generates the Standard Model neutrino masses. The model features an $SU(2)$ triplet with hypercharge $1$ alongside its twin counterpart. Spontaneous breaking of both $\mathbb{Z}_2$ and electroweak symmetry occurs in the scalar sector. The Standard Model neutrinos acquire small masses through the type-II seesaw mechanism. In contrast, their twin counterparts acquire large masses, effectively addressing the dark radiation problem in mirror twin Higgs scenarios. We study the impact of the model on the $N_{\rm eff.}$ constraints, as well as on collider phenomenology.}

\begin{document}
\maketitle
\flushbottom

\section{Introduction}
The standard model (SM) of particle physics is a successful quantum field theory describing the interactions of all known elementary particles. However, many questions are not addressed by it. Two of these are the stability of the electroweak (EW) scale and the origin of neutrino masses. The Higgs sector of the SM is quadratically sensitive to ultra-violet (UV) scales. Thus, if the SM cutoff were to be set high above the electroweak scale, the renormalization condition fixing the Higgs boson mass becomes fine tuned. A solution to the problem is to postulate that the SM cutoff, and therefore new physics, are not too far above the EW scale, typically not more than a loop factor above, i.e. at the TeV scale. This so-called  EW hierarchy problem is a motivation for introducing physics Beyond the Standard Model (BSM) at the $\TeV$ scale \cite{Craig:2022uua}. The traditional solutions to the hierarchy problem, such as Supersymmetry \cite{Carena:2002es} and composite Higgs models \cite{Panico:2015jxa}, predict the existence of new particles with masses of order $\TeV$, some of which are QCD color charged,  and would be responsible for taming the UV sensitivity from the SM contributions to the Higgs potential, making them good targets for collider resonance searches. 

On the other hand, since the discovery of neutrino oscillations \cite{ParticleDataGroup:2022pth,SNO:2002tuh,Super-Kamiokande:1998kpq}, it is clear that the SM neutrinos must have nonzero masses, which the SM cannot accommodate. For instance, the seesaw mechanism is a way to introduce new physics that would lead to neutrino masses, as well as to an explanation of their size compared to other fermion masses. There are three incarnations of the seesaw mechanism, type-I \cite{Minkowski:1977sc,Mohapatra:1979ia,Yanagida:1979as,Gell-Mann:1979vob,Glashow:1979nm,Schechter:1980gr}, type-II \cite{Schechter:1980gr,Mohapatra:1980yp,Lazarides:1980nt,Mandal:2022zmy}, and type-III \cite{Foot:1988aq}. The scale of the new physics can be as low as the $\TeV$ and as high as $10^{15} \GeV$, depending on the details. 

The search for new physics associated with the $\TeV$ scale at the large hadron collider (LHC) has not resulted in any discovery so far. In particular, states with color charges are severely constrained by the LHC data. Thus, an interesting possibility are neutral naturalness models as they provide an alternative solution to the stability of the EW scale that relies on having \textit{colorless partners} \cite{Batell:2022pzc,Craig:2022uua,Burdman:2006tz}. This is the case in twin Higgs models \cite{Chacko:2005pe,Chacko:2005vw,Barbieri:2005ri}, where the stability of the Higgs boson mass is achieved by the presence of colorless partners from a twin copy of the SM, supplemented by a discrete $\Z2$ symmetry. These new states are not subject to the strong constraints from top/gauge partner searches at the LHC \cite{Burdman:2014zta}. In the mirror twin Higgs model (MTH), the SM is extended with a mirror sector that contains a copy of all SM states and gauge symmetries. The Higgs doublet is a pseudo-Nambu-Goldstone boson (pNGB) of a spontaneously broken global symmetry. The $\Z2$ acts as an exchange symmetry between the SM and twin sectors, ensuring that the one loop corrections to the Higgs potential respect the original global symmetry, therefore not reintroducing the UV sensitivity of the Higgs boson mass present in the SM. 

The MTH model predicts deviations in the Higgs couplings to SM quarks and gauge bosons \cite{Burdman:2014zta}. The current experimental bounds from the measurements of the Higgs boson couplings at the LHC \cite{ATLAS:2021vrm,CMS:2022dwd} and the Higgs invisible decay width \cite{ATLAS:2023tkt} require the $\mathbb{Z}_2$ symmetry to be broken at IR scales. These measurements allow for a $10\%$ deviation in the Higgs coupling strength, meaning that the twin sector breaking scale $f$ should be somewhat higher than the EW vacuum expectation value (VEV), giving $f/v \gtrsim 3$. The $\mathbb{Z}_2$ symmetry is expected to be broken either explicitly \cite{Chacko:2005pe,Harigaya:2019shz}, spontaneously \cite{Beauchesne:2015lva,Beauchesne:2021opx}, or radiatively \cite{Yu:2016bku}. Furthermore, the original MTH is severely constrained by cosmology \cite{Chacko:2018vss,Chacko:2016hvu,Craig:2016lyx,Csaki:2017spo}. The presence of twin-neutrinos and the twin-photon in the original model results in new contributions to dark radiation at the Big Bang Nucleosynthesis (BBN) \cite{ParticleDataGroup:2022pth,Fields:2019pfx} and the Cosmic Microwave Background (CMB) \cite{Planck:2018vyg} epochs. These contributions are constrained by the observed effective number of relativistic degrees of freedom, with the most recent Planck measurement giving $N_{\rm eff} = 2.99_{-0.33}^{+0.34}$, \cite{Planck:2018vyg}, and from a recent combination of CMB with baryon acoustic oscillation (BAO) data from DESI~\cite{DESI:2024mwx}: $N_{\rm eff} = 3.10\pm 0.17$. 

There are several proposed solutions to the dark radiation problem in the cosmology of the MTH. One class of models proposes asymmetrically reheating the SM bath by some preferential decays to SM states \cite{Chacko:2016hvu,Craig:2016lyx}. 
In this work, we will focus on solutions that give large masses to twin neutrinos so that they do not contribute to $\Delta N_{eff}$. There are examples in the literature where large twin neutrino masses are achieved with soft $\Z2$ breaking and by introducing right-handed neutrinos. Then, using the type-I seesaw mechanism, one can reproduce light neutrino masses in the SM and heavy twin neutrinos \cite{Abada:2007sm,Bai:2015ztj,Batell:2015aha,Batell:2019ptb,Bishara:2018sgl,Holst:2023hff}. This method largely reduces the tension with dark radiation bounds. With the issue of dark radiation addressed, then it is possible to investigate other MTH features such as the nature of dark matter and mechanisms of baryogenesis \cite{Bittar:2023kdl,Farina:2015uea,Farina:2016ndq,Beauchesne:2020mih,Curtin:2021spx,Curtin:2021alk,Alonso-Alvarez:2023bat,Chacko:2021vin,GarciaGarcia:2015pnn,Holst:2023hff,Bodas:2024idn,Craig:2015xla,GarciaGarcia:2015fol,Cheng:2018vaj,Hochberg:2018vdo,Badziak:2019zys,Koren:2019iuv,Terning:2019hgj,Berbig:2022hsm,Ritter:2024sqv}.

In this paper we present an MTH model with spontaneous breaking of the $\mathbb{Z}_2$ symmetry, which results in large masses for the twin neutrinos, as well as in the small masses for the SM neutrinos. For this purpose, we add an $SU(2)$ triplet with hypercharge $1$, as well as its twin copy. The vacuum expectation value of the scalar sector breaks $\mathbb{Z}_2$ and electroweak symmetry spontaneously. Then, by the type-II seesaw mechanism, the SM neutrinos get small masses, while their twin counterparts get large masses controlled by the $\Z2$ breaking parameter, therefore mitigating the dark radiation problem. In Section \ref{sec:SSB}, we present the model and show the spontaneous $\mathbb{Z}_2$ symmetry breaking pattern. Then, we investigate the vacuum structure of the theory and its spectrum. We obtain neutrino masses and discuss their phenomenology in Section \ref{sec:neutrinomasses}, and discuss LHC bounds in Section \ref{sec:LHCpheno}. Other constraints, such as those from domain walls and the resulting twin Majoron from the spontaneous breaking of lepton number, are discussed in Section~\ref{sec:DomainWalls}. We conclude in Section~\ref{sec:conc}.

\section{Spontaneous \tstring{$\mathbb{Z}_2$}{Z2} breaking with the type-II Seesaw MTH}
\label{sec:SSB}

Following \cite{Chacko:2005pe}, we assume the MTH model where the Higgs is a (pseudo) Nambu-Goldstone boson (pNGB) of the global symmetry breaking $SU(4)\times U(1)\rightarrow SU(3)\times U(1)$. The SM Higgs doublet $H_A$ is part of a complex scalar bi-doublet
\begin{align}
\begin{gathered}
    H=\left( H_A, H_B\right)^T,
    \vspace{0.2cm} \\ 
    H_A=\left(G_A^+, \frac{h_A + i G_A^0}{\sqrt{2}}\right)^T,
    \hspace{1cm}
    H_B=\left(G_B^+, \frac{h_B + i G_B^0}{\sqrt{2}}\right)^T
\end{gathered}
\end{align}

\noindent which transforms under the fundamental representation of $SU(4)$. The indices $A, B$ designate the SM particles and twin partners, respectively. The global symmetry is broken explicitly by gauge and Yukawa interactions; however, due to the $\Z2$ symmetry, the effective potential is accidentally $SU(4)\times U(1)$ symmetric at the one loop level. We introduce additional $SU(2)_{A,B}$ triplets to the scalar sector of the theory, $\Delta_{A,B}$. The SM triplet has hypercharge $Y_\Delta = 1$, and the triplet partner has twin hypercharge $\widetilde{Y}_\Delta = 1$. We can write the triplet fields as
\begin{equation}
    \Delta_A=\frac{1}{\sqrt{2}}\left(\begin{array}{cc}
    \Delta_A^{+} & \sqrt{2}\Delta_A^{++}
    \\
    \sqrt{2}\Delta_A^0 & -\Delta_A^+
    \end{array}\right), \qquad
    \Delta_B=\frac{1}{\sqrt{2}}\left(\begin{array}{cc}
    \Delta_B^{+} & \sqrt{2}\Delta_B^{++}
    \\
    \sqrt{2}\Delta_B^0 & -\Delta_B^+
    \end{array}\right),
\end{equation}
where $\Delta_{A,B}^{++}, \Delta_{A,B}^{+}$ and $\Delta_{A,B}^{0}$ are the charge/twin charge eigenstates with $Q=(2,1,0)$ respectively. We can divide the neutral fields into the CP even, $\Delta_{A_s}^0, \Delta_{B_s}^0$, and CP odd, $\Delta_{A_a}^0, \Delta_{B_a}^0$, components as 
\begin{equation}
    \Delta_A^0 =\frac{1}{\sqrt{2}}( V + \Delta_{A_s}^0 + i \Delta_{A_a}^0)
    , \hspace{1cm} 
    \Delta_B^0 =\frac{1}{\sqrt{2}}( F + \Delta_{B_s}^0 + i \Delta_{B_a}^0),
\end{equation}
where we defined $V$ and $F$ as the triplets vacuum expectation values (VEVs).
\noindent Ignoring quarks, we will consider the following  Lagrangian respecting gauge and the $\mathbb{Z}_2$ symmetry:

\begin{align}
    \begin{split}
        -{\cal L} 
        &=-(DH_A)^\dagger DH_A - (DH_B)^\dagger DH_B + m^2 H^\dagger H + \lambda(H^\dagger H)^2  + \lambda' H^\dagger_A H_AH^\dagger_B H_B
        \\
        & -  (D\Delta_A) ^\dagger D\Delta_A - (D\Delta_B) ^\dagger D\Delta_B + M^2({\rm Tr}\Delta_A^\dagger \Delta_A +{\rm Tr}\Delta_B^\dagger \Delta_B) +\lambda_1\left({\rm Tr}\Delta_A^\dagger \Delta_A + \right. 
        \\ 
        & \left.+ {\rm Tr}\Delta_B^\dagger \Delta_B \right)^2 + \lambda_2\left({\rm Tr}\Delta_A^\dagger \Delta_A\right)\left({\rm Tr}\Delta_B^\dagger \Delta_B\right)  + \lambda_3\left({\rm Tr}\Delta_A^\dagger \Delta_A \Delta_A^\dagger \Delta_A +{\rm Tr}\Delta_B^\dagger \Delta_B  \Delta_B^\dagger\Delta_B\right)
        \\
        &+\lambda_4 \left(H_A^\dagger H_A+H_B^\dagger H_B\right)\left({\rm Tr}\Delta_A^\dagger \Delta_A + {\rm Tr}\Delta_B^\dagger \Delta_B \right) + \lambda_5\left(H_A^\dagger\Delta_A \Delta_A^\dagger H_A + H_B^\dagger\Delta_B \Delta_B^\dagger H_B\right)
        \\
        &+ \left\{b^*\left(H_A^\dagger\Delta_A \epsilon H_A^* + H_B^\dagger\Delta_B \epsilon H_B^*\right) + \kappa\left(\ell^T_A\epsilon\Delta_A\epsilon \ell_A + \ell^T_B\epsilon\Delta_B\epsilon \ell_B\right)+{\rm h.c.}\right\}.
    \end{split}
    \label{eq:lagrangian}
\end{align}

\noindent where $\epsilon_{12}=-\epsilon_{21}=1$.
In addition to the $SU(4)$ preserving quadratic term for the doublets with coefficient $m^2$, we have an $SU(4)$ preserving triplet mass term with coefficient $M^2$, as well as an $SU(4)$ breaking but $\Z2$ preserving quartic coupling $\lambda'$. In the third line in \eqref{eq:lagrangian} above, we have the triplet self-interactions. In the fourth line, we present the possible renormalizable and $\Z2$ preserving doublet-triplet interactions, where the first term is additionally $SU(4)$ preserving~\footnote{In fact, for (\ref{eq:lagrangian}) to be the most general Lagrangian respecting the gauge and $Z_2$ symmetries, the first term of the fourth line  should be split in two: $\lambda_4 \left( H_A^\dagger H_A\Tr
  \Delta_A^\dagger \Delta _A + H_B^\dagger H_B\Tr \Delta_B^\dagger
  \Delta_B\right) + \lambda_4' \left( H_A^\dagger H_A\Tr
  \Delta_B^\dagger \Delta _B + H_B^\dagger H_B\Tr \Delta_A^\dagger
  \Delta_A\right)$. But in (\ref{eq:lagrangian}) we took $\lambda_4=\lambda_4'$, which turns this into an $SU(4)$ respecting term. This is just a simplification, since we will use the $SU(4)$ breaking necessary to align the vacuum as desired in the following term. See discussion around Eqs. (\ref{eq:lam5term})-(\ref{eq:spontaneous_Z2})}. Finally, in the last line, the first term corresponds to the cubic doublet-triplet couplings. The last term in \eqref{eq:lagrangian} denotes the triplet interactions to the visible and twin left-handed leptons. The last two terms will be responsible for neutrino masses in both sectors. If we choose the dimensionful parameter $b$ to be small enough, it is possible to implement the type-II seesaw model for neutrino masses. The crucial point is that the spontaneous $\Z2$ breaking will result in a significant VEV $F$ for the twin triplet $\Delta_B$. In contrast, the visible sector triplet $\Delta_A$ has a practically vanishing VEV, $V$. If we can arrange the potential in \eqref{eq:lagrangian} to accommodate this situation, we would not only break the $\Z2$ symmetry spontaneously but also give large masses to the twin neutrinos. At the same time, by a judicious choice of a value of $b$, we can get small enough visible neutrino masses.    

We want to investigate the minimum of the potential associated with this model and show that this minimum can produce spontaneous $\mathbb{Z}_2$ and electroweak symmetry breaking. However, it is hard to find closed analytical formulas in full generality. It will be advantageous to start by studying the case in which $b$ is small compared with the other dimensional parameters of the model since, in this case, the potential greatly simplifies, and we can find simple expressions that hold to first order in $b$. 

Before analyzing the minima of the scalar potential in this limit, it is useful to consider a restricted set of operators to see how spontaneous $\Z2$ breaking can be obtained in the theory. For the twin Higgs quadratic terms, the $\Z2$ symmetry makes the global $SU(4)$ an accidental symmetry at the 1-loop level. However, $SU(4)$ is expected to be explicitly broken in the quartic terms by gauge interactions, the Yukawa couplings, and possibly other effects \cite{Chacko:2005vw}. 
Then, it is natural to write the $\Z2$-preserving and $SU(4)$-violating operator
\begin{equation}
\label{eq:lam5term}
    \lambda_5 \left(H_A^\dagger\Delta_A \Delta_A^\dagger H_A + H_B^\dagger\Delta_B \Delta_B^\dagger H_B\right).
\end{equation}
If we now assume that the scalar fields acquire VEVs along the neutral field directions, we can have
\begin{equation}
    |\langle H_A\rangle|^2=\frac{v^2}{2}, \qquad
    |\langle H_B\rangle|^2=\frac{f^2}{2}, \qquad
    |\langle \Delta_A\rangle|^2=\frac{V^2}{2}\sim 0, \qquad
    |\langle \Delta_B\rangle|^2=\frac{F^2}{2}.
    \label{eq:vevs}
\end{equation}

\noindent We are interested in a potential with the proper structure to allow for $V^2\sim 0$, as this vacuum structure can lead to small neutrino masses as we demonstrate in Section~\ref{sec:neutrinomasses}. Taking the vevs of the triplets as in \eqref{eq:vevs}, we obtain the effective $\Z2$ breaking term
\begin{equation}
    m^2 |H|^2 +\lambda_5 \left(H_A^\dagger\Delta_A \Delta_A^\dagger H_A + H_B^\dagger\Delta_B \Delta_B^\dagger H_B\right) \rightarrow  m^2 h_A^2  +\left(m^2+\lambda_5 F^2\right)  h_B^2,
    \label{eq:spontaneous_Z2}
\end{equation}
where $h_{A,B}$ are the radial modes of the doublets. Assuming that both $m^2$ and $\lambda_5$ are negative, the overall result is a net increase of the twin sector breaking scale compared to the SM, as desired. Thus, we see that including \eqref{eq:lam5term} in the Lagrangian~\eqref{eq:lagrangian}, among other $SU(4)$-preserving and -violating operators, will allow us to look for phases that have the appropriate vacuum structure we seek, i.e. that in \eqref{eq:vevs}. In principle, we could have also split the previous term in (\ref{eq:lagrangian}) (proportional to $\lambda_4$) into two $SU(4)$ breaking, $Z_2$ preserving terms, opening up even more parameter space to obtain the desired vacuum alignment. However, we opted for simplicity and left this term $SU(4)$ preserving. (See footnote in the previous page.).

Assuming $v\neq 0, f\neq 0$ and $F\neq 0$, we can expand the fields around their VEVs and minimize the potential to get
\begin{subequations}
\begin{alignat}{5}
   \label{eq:Vmin1}& -2 b V + m^2 + 2 (f^2 + v^2)\lambda + (F^2 +V^2) \lambda_4 + V^2 \lambda_5 + f^2 \lambda' && \,= 0, \\
   \label{eq:Vmin2} & - b v^2 + V \left( M^2+F^2(2\lambda_1 + \lambda_2)+2 V^2 \lambda_{13} + f^2 \lambda_4 +v^2 \lambda_{45} \right) && \, =0, \\
   \label{eq:Vmin3}& -2 b F + m^2 + 2 (f^2+v^2)\lambda + (F^2+V^2)\lambda_4 + F^2\lambda_5 + v^2 \lambda' && \,=0, \\ 
   \label{eq:Vmin4}& - b f^2 + F \left( M^2+V^2(2\lambda_1 + \lambda_2)+2 F^2 \lambda_{13} + v^2 \lambda_4 +f^2 \lambda_{45} \right) && \, =0,
\end{alignat}
\end{subequations}

\noindent where we defined $\lambda_{13}=\lambda_1+\lambda_3$ and $\lambda_{45}=\lambda_4+\lambda_5$. From \eqref{eq:Vmin2}, if $b = 0$, we can see that $V=0$ is an extremum of the potential. Keeping $b=0$, we obtain a relationship between the different VEVs by subtracting \eqref{eq:Vmin3} from \eqref{eq:Vmin1},
\begin{equation}
    F^2 = \frac{\lambda'}{\lambda_5}(f^2-v^2).
    \label{eq:VEVs_relation}
\end{equation}

\noindent The critical point with $V=0$ in the $b=0$ limit is
\begin{subequations}
\begin{align}
	&v^2=\frac{M^2 (\lambda'\lambda_{45}+2 \lambda \lambda_5)-m^2 (2\lambda' \lambda_{13}+\lambda_5\lambda_{45})}{2 \lambda  \left(4\lambda' \lambda_{13}+\lambda_5^2\right)+2 \lambda' (\lambda' \lambda_{13}-\lambda_4\lambda_{45})},\\
	&f^2=\frac{M^2(\lambda'\lambda_4-2\lambda\lambda_5)-m^2 (2\lambda' \lambda_{13}-\lambda_4\lambda_5))}{2 \lambda  \left(4\lambda' \lambda_{13}+\lambda_5^2\right)+2 \lambda' (\lambda' \lambda_{13}-\lambda_4\lambda_{45})},\\
	&F^2=\frac{\lambda' \left(m^2 (\lambda_4+\lambda_{45})-M^2 (4 \lambda +\lambda')\right)}{2 \lambda  \left(4\lambda' \lambda_{13}+\lambda_5^2\right)+2 \lambda' (\lambda' \lambda_{13}-\lambda_4\lambda_{45})}.
\end{align}
\label{eq:criticalpoint}
\end{subequations}

\begin{figure}
    \centering
    \includegraphics[scale=0.32]{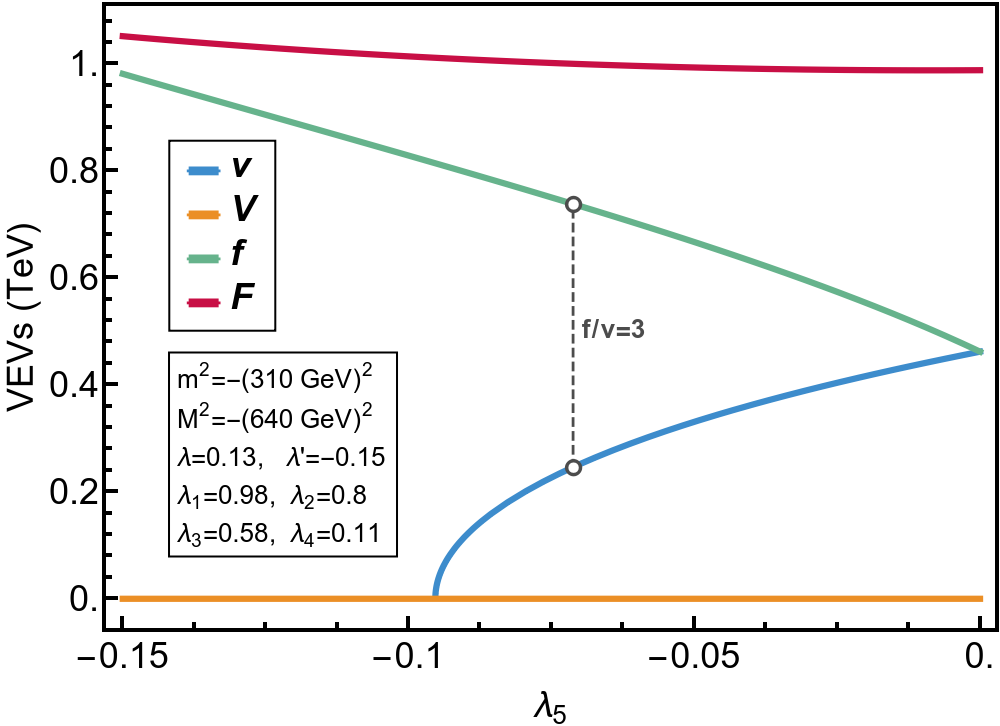}
    \caption{Vacuum expectation values of the neutral scalars in the model as a function of the $\lambda_5$ parameter of the potential \eqref{eq:lagrangian}. Through spontaneous $\Z2$ breaking, $\lambda_5$ controls the separation between the SM and twin sector. The VEVs for the doublets $H_A$ and $H_B$ are denoted as $v$ and $f$, while the VEVs for the triplets $\Delta_A$ and $\Delta_B$ are represented by $V$ and $F$ respectively. The parameters are chosen to have a realistic hierarchy of VEVs, with $f/v \approx 3$, as shown by the gray dashed line, and reproduce the correct SM Higgs mass.}
    \label{fig:vevs}
\end{figure}

\noindent As argued before, the spontaneous $\Z2$ breaking is controlled by the $\lambda_5$ parameter. Indeed, for $\lambda_5\rightarrow 0$, one can see from \eqref{eq:criticalpoint} that $v^2\rightarrow f^2$. Assuming negative $\lambda_5$, a nonzero value can generate the necessary hierarchy between the doublets, as shown in Figure \ref{fig:vevs}. The necessary hierarchy between the doublet VEVs, $f/v \gtrsim 3$, can be achieved for negative $\lambda_5$. Taking a negative $\lambda'$, from $\eqref{eq:VEVs_relation}$, the twin-triplet VEV can be kept larger than $f$ by having $\lambda'<\lambda_5$. In the figure, the parameters of the potential are fixed to have a realistic scalar spectrum, including the $125 \GeV$ SM Higgs and the scalar masses that are allowed by experimental constraints. These constraints are imposed in Section \ref{sec:LHCpheno} below.

When the triplet scalar coupling $b$ is nonzero, the VEV of $\Delta_A$ is non-vanishing and proportional to $b$. Assuming that $b$ is much smaller than the other scales of the theory, we have:
\begin{align}
V=b \left(\frac{v^2}{F^2(\lambda_2-2\lambda_3)+(v^2-f^2)\lambda_5} \right)+ \mathcal{O}(b^2/v)
\label{eq:VEV_tripletA}
\end{align}

\noindent A vacuum expectation value for the triplet contributes to the $T$ parameter that is constrained by electroweak precision measurements at LEP and the LHC. These constraints put a bound on the scale of the parameter $b$ as we discuss in Section~\ref{sec:LHCpheno}. 

One useful relation is obtained by inverting the expressions for $v^2$ and $f^2$ in \eqref{eq:vevs} to write $m$ and $M$,
\begin{subequations}
    \begin{align}
    & m^2=\frac{\lambda'\lambda_4-2\lambda\lambda_5}{2\lambda_5}v^2-\frac{\lambda'\lambda_{45}+2\lambda\lambda_5}{\lambda_5}f^2,
    \\
    & M^2=\frac{2\lambda_{13}\lambda'-\lambda_4\lambda_5}{2\lambda_5}v^2-\frac{2\lambda_{13}\lambda'+\lambda_5\lambda_{45}}{2\lambda_5}f^2.
    \end{align}
    \label{mM}
\end{subequations}

Now, we can compute the mass eigenvalues and analyze the spectrum of the model. The particle content of the model includes two CP-even, two CP-odd, and six charged $Q={\pm1,\pm2}$ fields for each sector. Due to electroweak and twin-electroweak breaking, the goldstone bosons get eaten by the six physical gauge and twin-gauge bosons. Thus, the physical fields can be divided as
\begin{itemize}
    \item CP-even: $h,H, \Delta_{A_s}^0, \delta_{B_s}^0$.
    \item CP-odd: $\Delta_{A_a}^0, J_B \,\,(\text{Majoron})$.
    \item Charged: $\Delta_A^\pm, \Delta_A^{\pm\pm}, \delta_B^{\pm}, \Delta_B^{\pm\pm}$.
    \item Goldstone bosons: $G_A^0, G_A^\pm, S_B^0, S_B^\pm$.
\end{itemize}
The fields $h,H,\delta_{B_s}^0,J_B,S_B^0,S_B^{\pm}$ and $ \delta_B^\pm$ are obtained after redefining the states from the interaction to the physical basis. The fields get rotated as $\{h_A, h_B, \Delta_{B_s}^0 \} \rightarrow \{h,H, \delta_{B_s}^0 \} $, $\{G_B^{\pm}, \Delta_B^\pm \} \rightarrow \{S_B^{\pm}, \delta_B^\pm \}$ and $\{G_B^0,\Delta_{B_a}^0 \} \rightarrow \{S_B^{0}, J_B \}$. Here, we neglect the mixing induced by the $b$ coupling in the CP-even block and $Q_B=\pm1$ block as the mixing angle is small. However, for the CP-odd twin block we must take the $b$ term into account since $J_B$ is a light Majoron as we describe below.

In the case of the triplet $\Delta_A$, the only mixing term appears due to the triple scalar coupling in \eqref{eq:lagrangian} and is proportional to $b$. The neutral triplet fields are divided between CP even and CP odd components as $\Delta_A^0 =\frac{1}{\sqrt{2}}( V + \Delta_{A_s}^0 + i \Delta_{A_a}^0)$. Since $b$ is much smaller than other scales of the theory, the mixing angle of $\Delta_{A_s}$ with $h_A$ is negligible, and the masses are 
\begin{subequations}
\begin{align}
    & m_{\Delta_A^{\pm\pm}}^2 = \frac{(\alpha+\lambda_5) f^2-\alpha v^2}{2},\quad \alpha\equiv\frac{\lambda'(\lambda_2-2 \lambda_3)}{\lambda_5}
    \\
    & m_{\Delta_A^\pm}^2 =m_{\Delta_A^{\pm\pm}}^2+\frac{\lambda_5}{4}v^2, 
    \\
    & m_{\Delta_{A_a}^0}^2=m_{\Delta_A^{\pm\pm}}^2+\frac{\lambda_5}{2} v^2, 
    \\
    & m_{\Delta_{A_s}^0}^2 =m_{\Delta_{A_a}^0}^2.
\end{align}
\label{eq:masses_tripletsA}
\end{subequations}

\noindent In the twin sector, the $\Delta_B^{\pm\pm}$ component of the twin triplet does not get mixed with other scalar components. This is also true when $b=0$ for the $\Delta_{B_a}^0$ field. Therefore, their masses are obtained simply by the second derivative of the potential,
\begin{align}
    & m_{\Delta_B^{\pm\pm}}^2=-\frac{2 \lambda' \lambda_3+\lambda_5^2}{2\lambda_5} f^2+\frac{\lambda' \lambda_3}{\lambda_5}v^2,
    \\
    & m_{\Delta_{B_a}^0}^2=0 \quad (\text{if $b=0$}).
\end{align}

\noindent While the doubly charged scalar has a mass of order $\TeV$, the massless pseudoscalar twin state $\Delta_{B_a}^0$ is a Majoron - the nambu goldstone (NGB) boson of spontaneously broken lepton number. In the absence of the $b$ coupling, lepton number is conserved in the theory as $\Delta_A,\Delta_{B}$ have lepton/twin-lepton number $L=2$ and $\widetilde L =2$ respectively. When $ \Delta_B$ acquires a VEV, twin lepton number is spontaneously broken and a twin Majoron is generated only in the twin sector. When the triple scalar coupling is nonzero, the twin pseudoscalars $\{G_B^0,\Delta_{B_a}^0 \}$ mix, leading to the mass matrix,
\begin{equation}
    \mathcal{M}^2(\{G_B^0,\Delta_{B_a}^0 \}) = 
    \begin{pmatrix}
        2\sqrt{2} b F & -\sqrt{2}b f \\
        -\sqrt{2} b f & \frac{b f^2}{\sqrt{2}F}
    \end{pmatrix}.
\end{equation}

\noindent We can rewrite the fields in the physical basis 
\begin{align}
	& S_{B}^{0}=\cos\alpha \, G_B^0  + \sin\alpha \, \Delta_{B_a}^0,\\
	& J_{B}=\cos\alpha \, \Delta_{B_a}^0  - \sin\alpha \, G_B^0,
\end{align}

\noindent to find the masses of $S_B^0$ and $J_B$ and the mixing angle $\alpha$,
\begin{align}
    & m_{S_B^0}^2=0 \\
    &\label{eq:majoron_mass} m_{J_B}^2=b\frac{(f^2+4 F^2)}{\sqrt{2} F}\\
    & \label{eq:majoron_alpha} \sin 2\alpha = -\frac{f F}{f^2+4 F^2}~,
\end{align}
where the $S_B^0$ is the NGB eaten by the twin gauge sector, and $J_B$ is the pseudo-NGB associated with the spontaneous breaking of lepton number, the physical Majoron.
\noindent  We discuss the phenomenological consequences of the twin Majoron in Section~\ref{sec:pheno}.

The mass matrix of the CP even fields $\{ h_A, h_B, \Delta_{B_s}^0\}$ is given by
\begin{align}
    \mathcal{M}^2({\{ h_A, h_B, \Delta_{B_s}^0\}})=
    \begin{pmatrix}
        2 \lambda  v^2 
        & (2 \lambda +\lambda') v f 
        & (\lambda_4 \eta) v f 
        \\
        (2 \lambda +\lambda') v f  
        & 2 f^2 \lambda  
        &(\lambda_{45} \eta) f^2
        \\
        (\lambda_4 \eta) v f  
        & (\lambda_{45}\eta) f^2
        & 2\lambda_{13}\eta^2 f^2 
    \end{pmatrix},
    \label{eq:M2_CPeven}
\end{align}

\noindent where we have defined $\eta^2\equiv \frac{\lambda'(1-\xi)}{\lambda_5} = \frac{F^2}{f^2} + \mathcal{O}(b^2/f^2)$ and $\xi\equiv \frac{v^2}{f^2}$. In general, the expressions for the mass eigenvalues are complicated and not very illuminating. However, the mixing matrix is constrained by the Higgs invisible decay width measurements, as we discuss in Section~\ref{sec:pheno}. To limit the mixing of the Higgs, we need to tune the quartic couplings that control the off-diagonal terms of \eqref{eq:M2_CPeven},
\begin{align}
    2 \lambda +\lambda' \lesssim \frac{v^2}{f^2}, \hspace{1cm} \lambda_4 \eta \lesssim \frac{v^2}{f^2}.
    \label{eq:tuning}
\end{align}

\noindent After tuning the couplings as in \eqref{eq:tuning}, the SM Higgs is aligned with the $h_A$ state up to corrections of order $v/f$. Then, we can redefine $h_B$ and $\Delta_{B_s}^0$ as the physical states $H$ and $\Delta_{B_s}^0$ by
\begin{align}
    & H=\cos\theta \, h_B +\sin\theta \, \Delta_{B_s}^0,
    \\
    & \delta_{B_s}^0=\cos\theta \Delta_{B_s}^0 -\sin\theta h_B,
\end{align}

\noindent Therefore, up to order $v/f$, we have the following masses and mixing angle $\theta$,
\begin{align}
    & m_h^2 = 2 \lambda v^2 + \mathcal{O}(v^2 \sqrt{\xi})
    \\
    & m_H^2 = f^2 \left[ (\lambda + \eta^2 \lambda_{13}) - \sqrt{(\lambda - \eta^2 \lambda_{13})^2 + \eta^2 \lambda_5^2}\right] + \mathcal{O}(v^2 \sqrt{\xi})
    \\
    & m_{\delta_{B_s}^0}^2 = f^2 \left[ (\lambda + \eta^2 \lambda_{13}) + \sqrt{(\lambda - \eta^2 \lambda_{13})^2 + \eta^2 \lambda_5^2}\right] + \mathcal{O}(v^2 \sqrt{\xi}) 
    \\
    &\sin 2\theta = \frac{\eta \lambda_5}{\sqrt{(\lambda-\eta^2\lambda_{13})^2+\eta^2 \lambda_5^2}} + \mathcal{O}(v^2 \sqrt{\xi}).
\end{align}

\noindent The mass matrix with mixing fields $G_B^\pm$ and $\Delta_{B}^\pm$ is

\begin{equation}
    \mathcal{M}^2({\{ G_B^\pm,\Delta_{B}^\pm\}})=
    \begin{pmatrix}
    -\frac{1}{2}  \lambda_5 f^2 & \lambda_5 \eta v f
    \\
    \lambda_5 \eta v f
    & \hspace{0.2cm}\lambda' f^2\left(1-\xi\right)
\end{pmatrix}
\end{equation}

\noindent Now, we can rotate the fields to the mass basis,
\begin{align}
	& S_{B}^{\pm}=\cos\varphi \, G_B^\pm +\sin\varphi \, \Delta_B^\pm,\\
	& \delta_{B}^\pm=\cos\varphi \, \Delta_B^\pm -\sin\varphi \, G_B^\pm,
\end{align}

\noindent to find the mass eigenvalues and mixing angle
\begin{align}
    &m_{S_B^\pm}^2 = 0, \\
    &m_{\delta^\pm}^2 = -\frac{\lambda_5+2\lambda'}{4}f^2 +\frac{\lambda' }{2}v^2, \\
    &\sin 2\varphi = \frac{2\sqrt{2}\lambda_5 \eta f^2}{(\lambda_5+2\lambda')f^2-2\lambda' v^2}.
\end{align}

\noindent Finally, the masses of the eaten SM and neutral twin Goldstone bosons are zero, as expected.
\begin{equation}
    m_{G_A^0}^2 = m_{G_A^\pm}^2 = m_{G_B^0}^2 = 0.
\end{equation}

We conclude this section by commenting on how realistic the model is regarding the range of possible parameters and how commonly occurring the symmetry-breaking behavior described above is. As discussed above, some parameters must be tuned so that the CP even states do not mix too much. These conditions are given by \eqref{eq:tuning}. Because of this, the quartic coupling $\lambda$ should be close to the SM value up to order $v/f$. Additionally, both $\lambda'$ and $\lambda_5$ must be negative to obtain a lighter SM VEV. The model also requires that the dimensionfull parameter b should be small so that the SM triplet VEV is well below the EW scale, $V\ll v$. However, since setting b to zero restores lepton number symmetry in the theory, we see that a small value for b is still technically natural~\cite{tHooft:1979rat}, and radiatively stable. Given these conditions, it is straightforward to obtain a realistic spectrum of scalar and neutrino masses compatible with all constraints on the model. Table~\ref{tab:pars} shows one benchmark point we use throughout the paper. 

\begin{table}
    \centering
    \begin{tabular}{|c|c||c|c|}
        \hline
        Parameter & Value & \,\,\, Particle \,\,\, & \,\, Mass \,\, \\
        \hline
        $m^2$ & $-(310 \, \mathrm{GeV})^2$ & $m_{\nu_A}$ & $\sim 0.12 \, \mathrm{eV}$ \\
        $M^2$ & $-(640 \, \mathrm{GeV})^2$ & $m_{J_B}$ & $180 \, \mathrm{MeV}$ \\
        $\lambda$ & $0.13$ & $m_{\nu_B}$ & $280 \, \MeV$ \\
        $\lambda'$ & $-0.15$ & $m_h$ & $125 \, \mathrm{GeV}$ \\
        $\lambda _1$ & $0.98$ & $m_{\delta_B^{\pm}}$ & $370 \, \mathrm{GeV}$ \\
        $\lambda _2$ & $0.8$ & $m_H$ & $370 \, \mathrm{GeV}$ \\
        $\lambda _3$ & $-0.58$ & $m_{\Delta ^{\pm \pm }_{B_s}}$ & $775 \, \mathrm{GeV}$ \\
        $\lambda _4$ & $0.11$ & $m_{\delta ^0_{B_s}}$ & $894 \, \mathrm{GeV}$ \\
        $\lambda _5$ & $-0.071$ & $m_{\Delta ^0_{A_{s,a}}}$ & $1 \, \TeV$ \\
        $b$ & $10 \, \mathrm{keV}$ & $m_{\Delta ^{\pm }_{A_s}}$ & $\sim 1 \, \TeV$ \\
        $\kappa$ & $\sim 4\cross 10^{-4}$ & $m_{\Delta ^{\pm \pm }_{A_s}}$ & $\sim 1 \, \TeV$ \\
        \hline
    \end{tabular}
    \caption{Benchmark point chosen that satisfy all the constraints. On the left, we show the values of the parameters of the potential, and on the right, the resulting mass spectrum of the model.}
    \label{tab:pars}
\end{table}

Just as all other twin Higgs models, this one solves the little hierarchy problem, but it also maintains (almost ) exactly\footnote{We will need to introduce a very small explicit breaking of the $\Z2$ symmetry in order to avoid a domain wall problem. This is addressed in Section~\ref{sec:DomainWalls}.} the $\Z2$ symmetry in the potential. Additionally, including triplets with couplings to neutrinos equips us with the means to address the neutrino mass problem through the type-II seesaw model. In the next section, we show how the light SM neutrino masses can be obtained and how the twin neutrinos are kept heavy as a consequence of having a large twin triplet VEV. Then, we proceed to explore other phenomenological aspects of the model.

\section{Phenomenology}
\label{sec:pheno}

In this section, we examine the phenomenological implications of the model. We consider a combination of neutrino and collider phenomenology from triplet scalars in type-II seesaw models alongside novel features within the twin sector, including the presence of the light twin Majoron impacting cosmology. Finally, we discuss the formation of a domain wall network and show that it is very simple to implement a way of avoiding this problem.

\subsection{Neutrino and Twin Neutrino Masses}
\label{sec:neutrinomasses}
\begin{figure}
    \centering
    \includegraphics[scale=0.09]{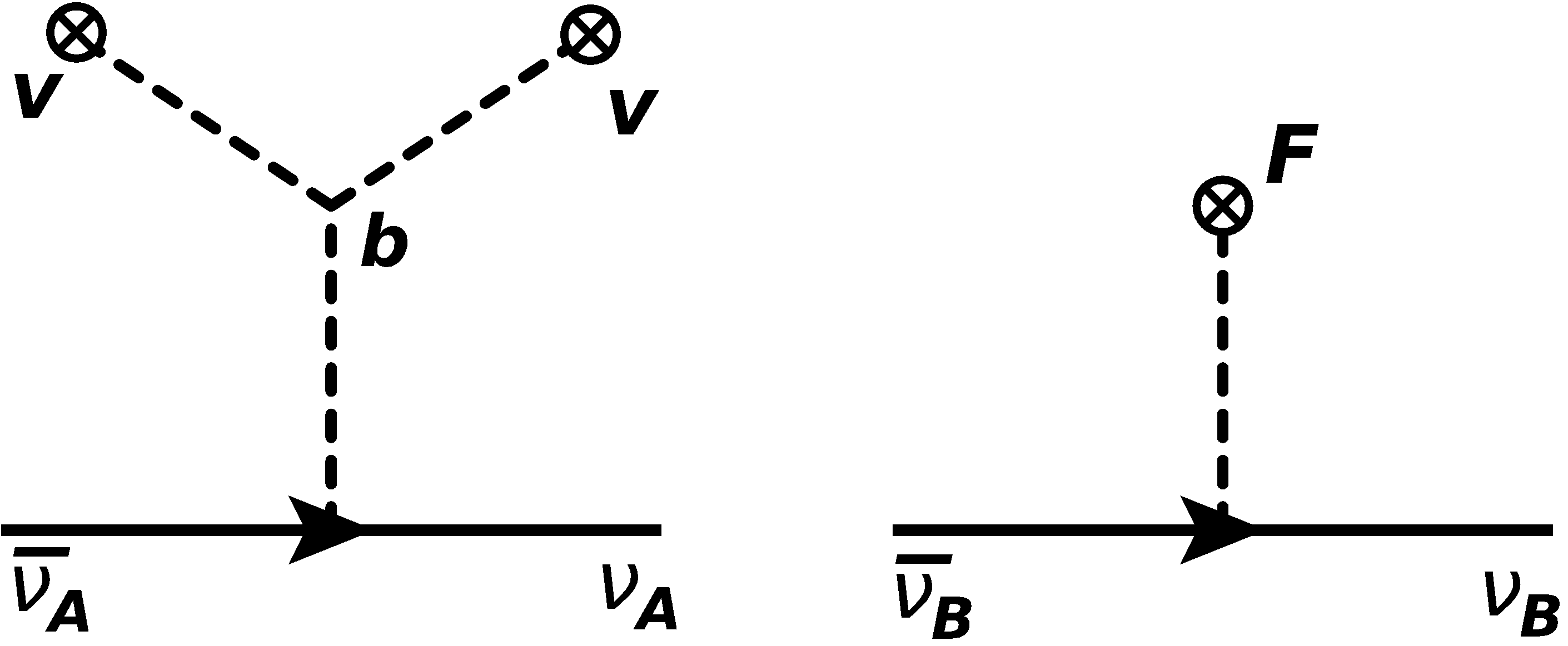}
    \caption{Neutrino mass generation from the interaction with the triplets. On the SM side (diagram on the left), the type-II seesaw mechanism is implemented. The SM neutrinos, $\nu_A$, are much lighter than the twin neutrinos since, in addition to the triplet mass, there is an extra suppression factor coming  from the $b$ coupling. On the twin neutrino side, the VEV of the triplet is much larger and results in a large twin neutrino mass directly (diagram on the right).}
    \label{fig:nu_masses}
\end{figure}

As we have seen, the twin triplet $\Delta_B$ acquires a large vacuum expectation value of the order of the $\TeV$ scale, while $\Delta_A$ has a much smaller VEV proportional to the $b$ coupling. As shown in Figure \ref{fig:nu_masses}, this hierarchy allows us to have large twin-neutrino masses and small SM neutrino masses. After EW and $\mathbb{Z}_2$ spontaneous symmetry breaking we have
\begin{align}
    \mathcal{L} & \supset -\frac{V}{\sqrt{2}} \kappa_{ij} \overline{\nu_A^c}_{,i} \nu_{A,j} -\frac{F}{\sqrt{2}} \kappa_{ij} \overline{\nu_B^c}_{,i} \nu_{B,j}~,
\end{align}
where $i,j$ are flavor indices. 

\noindent Therefore, we can write the following neutrino and twin-neutrino mass matrices,
\begin{align}
    & M_{\nu_A}= \sqrt{2}\kappa V = \sqrt{2} \kappa b  \frac{v^2}{m_\Delta^2} ,
    \label{eq:SM_nu_mass}
    \\ & M_{\nu_B} = \sqrt{2}\kappa F = \sqrt{2} \kappa f \sqrt{\frac{\lambda'}{\lambda_5}(1-\xi)}.
	\label{neutrinomasses}
\end{align}

\noindent where $m_\Delta^2 \equiv F^2(\lambda_2 - 2\lambda_3) + (v^2-f^2)\lambda_5$.

For the SM neutrinos, we can express the free parameters in $\kappa$ as functions of the neutrino oscillation parameters and the triplet VEV,
\begin{align}
    \kappa = \frac{1}{2V}(U_{PMNS})^* m_\nu^{\rm diag} (U_{PMNS})
\end{align}

\noindent The PMNS matrix in terms of the oscillation parameters and Majorana phases is given by

\begin{equation}
    U_{\text{PMNS}} = 
    \begin{pmatrix}
        c_{12}c_{13} & s_{12}c_{13} & s_{13}e^{-i\delta} 
        \\
        -s_{12}c_{23} - c_{12}s_{23}s_{13}e^{i\delta} & c_{12}c_{23} - s_{12}s_{23}s_{13}e^{i\delta} & s_{23}c_{13} 
        \\
        s_{12}s_{23} - c_{12}c_{23}s_{13}e^{i\delta} & -c_{12}s_{23} - s_{12}c_{23}s_{13}e^{i\delta} & c_{23}c_{13}
    \end{pmatrix}
    \begin{pmatrix}
        e^{i\alpha_1/2} & 0 & 0 
        \\
        0 & e^{i\alpha_2/2} & 0 
        \\
        0 & 0 & 1
    \end{pmatrix}
\end{equation}

\noindent with $c_{ij} = \cos{\theta_{ij}}$, $s_{ij} = \sin{\theta_{ij}}$, and $\delta$, $\alpha_1$, and $\alpha_2$ are the CP-violating phase and Majorana phases, respectively. For the \textit{Normal Hierachy} (NH) of neutrino masses, $m_1\ll m_2 < m_3$, the full set of oscillation parameters is
\begin{align}
    &\text{NH:} \quad \{m_{\nu_1},\Delta m^2_{21},\Delta m^2_{31},\theta_{12}, \theta_{13}, \theta_{23},\delta,\alpha_1,\alpha_2  \},
    \\
    & m_{\nu_2}=\sqrt{m_{\nu_1}^2 + \Delta m^2_{21}}, \qquad m_{\nu_3}=\sqrt{m_{\nu_1}^2 + \Delta m^2_{31}}.
\end{align}

\noindent Conversely, the parameters of the \textit{Inverted Hierachy} (IH), $m_3\ll m_2 < m_1$, are
\begin{align}
    &\text{IH:} \quad \{m_{\nu_3},\Delta m^2_{21},\Delta m^2_{32},\theta_{12}, \theta_{13}, \theta_{23},\delta,\alpha_1,\alpha_2  \},
    \\
    & m_{\nu_1}=\sqrt{m_{\nu_3}^2 - (\Delta m^2_{32} + \Delta m^2_{21})}, \qquad m_{\nu_2}=\sqrt{m_{\nu_3}^2 - \Delta m^2_{32}}.
\end{align}

\noindent Where, $\Delta m^2_{ij} = m^2_i -m^2_j$.These parameters can be determined from the combined analysis of the latest neutrino oscillation data as shown in Table \ref{tab:NuFIT}. Even with complete knowledge of the oscillation parameters, oscillation data does not fix the overall neutrino scale. To estimate the effective scale of neutrino masses, we can use the upper bound value from Planck data from fitting the $\Lambda$CDM model with a free sum of neutrino masses parameter, $\sum m_\nu$, \cite{Planck:2018vyg,ParticleDataGroup:2022pth},
\begin{align}
   \sum m_\nu < 0.12 \eV.
\end{align}

\begin{table}\centering
  \begin{footnotesize}
    \begin{tabular}{|l|cc|cc|}
      \hline
      & \multicolumn{2}{c|}{Normal Ordering (best fit)}
      & \multicolumn{2}{c|}{Inverted Ordering ($\Delta\chi^2=2.7$)}
      \\
      \hline
      & bfp $\pm 1\sigma$ & $3\sigma$ range
      & bfp $\pm 1\sigma$ & $3\sigma$ range
      \\
      \hline
      \rule{0pt}{4mm}\ignorespaces
        $\sin^2\theta_{12}$
      & $0.304_{-0.012}^{+0.013}$ & $0.269 \to 0.343$
      & $0.304_{-0.012}^{+0.013}$ & $0.269 \to 0.343$
      \\[1mm]
        $\sin^2\theta_{23}$
      & $0.570_{-0.024}^{+0.018}$ & $0.407 \to 0.618$
      & $0.575_{-0.021}^{+0.017}$ & $0.411 \to 0.621$
      \\[1mm]
        $\sin^2\theta_{13}$
      & $0.02221_{-0.00062}^{+0.00068}$ & $0.02034 \to 0.02430$
      & $0.02240_{-0.00062}^{+0.00062}$ & $0.02053 \to 0.02436$
      \\[1mm]
        $\delta_{CP}/^\circ$
      & $195_{-25}^{+51}$ & $107 \to 403$
      & $286_{-32}^{+27}$ & $192 \to 360$
      \\[1mm]
        $\dfrac{\Delta m^2_{21}}{10^{-5} \eV^2}$
      & $7.42_{-0.20}^{+0.21}$ & $6.82 \to 8.04$
      & $7.42_{-0.20}^{+0.21}$ & $6.82 \to 8.04$
      \\[2mm]
        $\dfrac{\Delta m^2_{3l}}{10^{-3} \eV^2}$
      & $+2.514_{-0.027}^{+0.028}$ & $+2.431 \to +2.598$
      & $-2.497_{-0.028}^{+0.028}$ & $-2.583 \to -2.412$
      \\[2mm]
      \hline
    \end{tabular}
  \end{footnotesize}
  \caption{Neutrino oscillation parameters taken from the most recent global analysis from the \textit{NuFit} collaboration \cite{Esteban:2020cvm}. The results are shown without adding the Super Kamiokande atmospheric data.}
  \label{tab:NuFIT}
\end{table}

\noindent To get a small scale of neutrino masses in \eqref{eq:SM_nu_mass}, we need the combination of the $\kappa$ parameter and $V$ to be of order of $\eV$. Using \eqref{eq:VEV_tripletA} in \eqref{eq:SM_nu_mass}, we can parametrically write,
\begin{equation}
    b \simeq 10 \text{ keV} \left(\frac{m_\Delta}{1\TeV}\right)\left( \frac{\kappa_{max}}{4\times 10^{-4}}\right)
\end{equation}

\noindent As a benchmark point, we choose the largest eigenvalue of the coupling matrix, $\kappa_{max}$, to be of order $10^{-4}$, which implies $b\sim 10 \text{ keV}$. As we discuss in Section~\ref{sec:DomainWalls}, this choice minimizes the effective number of relativistic degrees of freedom as both the Majoron and the twin neutrinos are non-relativistic at decoupling.

We can also constrain the off-diagonal entries of $\kappa$ and the minimum value of $V$ by using data on flavor violation tests \cite{Ilakovac:1994kj,Barrie:2022ake,Cheng:2022jyi,Alves:2022yav,Alves:2023znq}. Following \cite{Cheng:2022jyi}, we can calculate the effective interaction due to the exchange of a heavy $\Delta_A^{\pm\pm}$ particle,
\begin{equation}
    \mathcal{L}_{\rm eff} (\Delta_A^{\pm\pm})= \frac{(M_{\nu_A})_{ij}(M_{\nu_A}^*)_{kl}}{4m_\Delta^2 V^2} \left[ \overline{e_{L}}_{,l} \gamma^\mu e_{L,j}\right] \left[ \overline{e_{L}}_{,k} \gamma^\mu e_{L,i} \right].
\end{equation}

\noindent Then, we can calculate the branching ratio of the processes $l_i^- \rightarrow l_j^+ l_k^- l_l^-$.
\begin{equation}
    \text{BR}(l_i^- \rightarrow l_j^+ l_k^- l_l^-) = \frac{1}{8(1+\delta_{kl})}\frac{m_{l_i}^5 \tau_i}{192\pi^3}\left| \frac{(M_{\nu_A})_{ij}(M_{\nu_A}^*)_{kl}}{m_\Delta^2 V^2}\right|^2, 
\end{equation}

\noindent where $\tau_i$ is the lifetime of $l_i$. Besides the flavor violating decays to $3$ leptons, muon and tau decays to electrons and photons are also important and can constrain the parameters of the model. The loop induced branching ratio of $l_i^- \rightarrow l_j^- \gamma$ due to exchange of intermediate $\Delta^{+}$ and $\Delta^{++}$ is given by
\begin{equation}
    \text{BR}(l_i^- \rightarrow l_j^- \gamma) = \frac{m_{l_i}^5 \alpha_{\rm em} \tau_i}{(192\pi^2)^2}\left( \frac{9|(M_{\nu_A}^\dagger M_{\nu_A})_{ji}|}{2m_\Delta^2 V^2}\right)^2
\end{equation}

\noindent In Table~\ref{tab:lfv}, we list the lepton flavor violation bounds for different combinations of the couplings $\kappa$. The bounds on the squared matrix $\kappa^\dagger \kappa$ can be converted into a lower bound on the scale of $V$ by using the following relation
\begin{equation}
    \left| (\kappa^\dagger \kappa)_{ij}\right| = \frac{1}{2V}\left| U^{i2}_{\rm PMNS} U^{j2*}_{\rm PMNS} \Delta m_{21}^2 + U^{i3}_{\rm PMNS} U^{j3*}_{\rm PMNS} \Delta m_{31}^2 \right|.
    \label{eq:kappa_sq}
\end{equation}

\begin{table}
	\begin{center}
		\begin{tabular}{|l|c|l|}
		\hline
            Process & Branching ratio bound &  \multicolumn{1}{c|} {Constraint}
            \\\hline        
            $\mu^- \to e^+e^-e^-$ 
            & $1.0\times 10^{-12}$ 
            & $\left|\kappa^{*}_{e e} \kappa_{\mu e}\right| < 2.36 \times 10^{-5}\left(\frac{m_{\Delta}}{1 \TeV }\right)^{2}$
            \\
            $\tau^- \to e^+e^-e^-$ 
            & $2.7\times 10^{-8}$ 
            & $\left|\kappa^{*}_{e e}\kappa_{\tau e}\right| < 9.08 \times 10^{-3}\left(\frac{m_{\Delta}}{1 \TeV}\right)^{2}$ 
            \\
            $\tau^- \to e^+e^-\mu^-$ 
            & $1.8\times 10^{-8}$ 
            & $\left|\kappa^{*}_{e \mu}\kappa_{\tau e}\right| < 5.24 \times 10^{-3}\left(\frac{m_{\Delta}}{1 \TeV}\right)^{2}$
            \\
            $\tau^- \to e^+\mu^-\mu^-$ 
            & $1.7\times10^{-8}$ 
            & $\left|\kappa^{*}_{\mu \mu}\kappa_{\tau e}\right| < 7.21 \times 10^{-3}\left(\frac{m_{\Delta}}{1 \TeV}\right)^{2}$ 
            \\
            $\tau^- \to \mu^+e^-e^-$ 
            & $1.5\times 10^{-8}$ 
            &$\left|\kappa^{*}_{e e}\kappa_{\tau \mu}\right| < 6.77 \times 10^{-3}\left(\frac{m_{\Delta}}{1 \TeV}\right)^{2}$ 
            \\
            $\tau^- \to \mu^+\mu^-e^-$ 
            & $2.7\times 10^{-8}$ 
            & $\left|\kappa^{*}_{\mu e} \kappa_{\tau \mu}\right| < 6.42 \times 10^{-3}\left(\frac{m_{\Delta}}{1 \TeV}\right)^{2}$ 
            \\
            $\tau^- \to \mu^+\mu^-\mu^-$ 
            & $2.1\times 10^{-8}$ 
            & $\left|\kappa^{*}_{\mu \mu}\kappa_{\tau \mu}\right| < 8.01 \times 10^{-3}\left(\frac{m_{\Delta}}{1 \TeV}\right)^{2}$
            \\[1mm]
            $\mu\to e\gamma$ 
            & $4.2\times 10^{-13}$ 
            & $\left|\left(\kappa^{\dagger} \kappa\right)_{e \mu}\right| <2.36 \times 10^{-4}\left(\frac{m_{\Delta}}{1 \TeV}\right)^{2}$
            \\[2mm]
		$\tau\to e\gamma$ 
            & $3.3\times 10^{-8}$ 
            & $\left|\left(\kappa^{\dagger} \kappa\right)_{e        \tau}\right| <1.55 \times 10^{-1}\left(\frac{m_{\Delta}}{1 \TeV}\right)^{2}$ 
            \\[1mm]
            $\tau\to \mu\gamma$ 
            & $4.4\times 10^{-8}$ 
            & $\left|\left(\kappa^{\dagger} \kappa\right)_{\mu \tau}\right| <1.79 \times 10^{-1}\left(\frac{m_{\Delta}}{1 \TeV}\right)^{2}$
            \\[1mm]
		\hline
	\end{tabular}
        \caption{Lepton flavor violation decay constraints for $|\kappa_{i j} \kappa^{*}_{k l}|$, $\left| \kappa^{\dagger} \kappa\right|_{j i}$ and $V$. Table adapted from \cite{Cheng:2022jyi}. Obtaining a lower bound for $V$ involves numerically searching for the minimum value of \eqref{eq:kappa_sq} within the $3 \sigma$ range for the mixing parameters presented in Table \ref{tab:NuFIT}.}
	\label{tab:lfv}
	\end{center}
\end{table}

\noindent Then, using the oscillation parameters from Table~\ref{tab:NuFIT} and the bounds from Table~\ref{tab:lfv}, one can numerically search for the minimum value of \eqref{eq:kappa_sq} in the $3 \sigma$ confidence limit and solve the inequality for $V$. The results for NH and IH are approximately the same as argued in \cite{Cheng:2022jyi}. Using the data from $\mu\rightarrow e \gamma$ we obtain the most stringent bound for $V$ as 
\begin{equation}
    V > 0.625 ~\mathrm{eV}\left(\frac{1 \TeV}{m_{\Delta}}\right).
\end{equation}

\noindent Due to the number of free parameters in $\kappa$, we can reproduce all of the necessary constraints from neutrino phenomenology. Because of this, we do not consider an in-depth analysis of the possible neutrino textures for $\kappa$ and leave the discussion for other papers that focus on type-II seesaw models more specifically \cite{Abada:2007ux,ParticleDataGroup:2022pth}.

Regarding the twin neutrinos, their masses are much larger than SM neutrinos. Large twin neutrinos alleviate the excess of radiation during BBN and in the CMB in the cosmological history. The SM 
and twin sectors are kept in equilibrium by the Higgs portal interactions up to temperatures around $T_{\rm D} \sim 3 \GeV$. As discussed in \cite{Chacko:2016hvu}, if the Higgs portal sets the decoupling temperature between the two sectors, the effective number of relativistic degrees of freedom in the early universe would be much above the present bounds from Planck \cite{Planck:2018vyg}. As in \cite{Farina:2015uea}, we assume the best-case scenario where the decoupling temperature is between the two QCD phase transitions. Twin QCD would already have condensed at temperatures around $\sim 200 \MeV$ while the SM quarks and gluons are still relativistic degrees of freedom.\footnote{A lower decoupling temperature can be achieved by including extra portals to the twin sector, i.e., kinetic mixing or effective operators at the twin Higgs cutoff.} Then, assuming the benchmark in Table~\ref{tab:pars}, the twin neutrinos are heavy and already non-relativistic at BBN and CMB times as well as when the two sectors decouple. We define the number of effective relativistic degrees of freedom from the energy density in radiation during BBN,
\begin{align}
    \rho_{\rm rad} = \rho_{\gamma_A} + \rho_{\nu_A} + \rho_{\gamma_B} \equiv \left[ 1+\frac{7}{8} N_{\rm eff} \left(\frac{T_{\nu_A}}{T_{\gamma_A}}\right)^4\right]\rho_{\gamma_A},
\end{align}

\noindent where $T_{\nu_A}$ and $T_{\gamma_A}$ are the temperatures of SM neutrinos and photons respectively. Using that $\frac{T_{\nu_A}}{T_{\gamma_A}}=\left(\frac{4}{11}\right)^{1/3}$, we can obtain the number of extra degrees of freedom by subtracting the SM contribution, $N_{\rm eff}^{\rm SM}\simeq 3.046$.

\begin{equation}
    \Delta N_{\rm eff}=N_{\rm eff}-N_{\rm eff}^{\rm SM}=\left[\frac{7}{8} \left(\frac{4}{11}\right)^{4/3}\right]^{-1}\left( \frac{T_{\gamma_B}}{T_{\nu_A}}\right)^4 .
    \label{eq:DeltaNeff_def}
\end{equation}

\noindent Finally, using conservation of entropy, we can calculate the ratio of the temperatures of twin photons by SM neutrinos as $\frac{T_{\gamma_B}}{T_{\nu_A}}= \left( \frac{g_{*S}^B(T_{\rm dec})}{g_{*S}^A(T_{\rm dec})}\right)^{1/3}$. At decoupling between the SM and twin sectors, the latter contains only the twin photon and twin electrons. Conversely, the SM has many more degrees of freedom, with $g_{*S}^A(T_{\rm dec})\approx 61.75$ after $c\overline{c}$ annihilation. Therefore, we have
\begin{equation}
    \Delta N_{\rm eff}=\left[\frac{7}{8} \left(\frac{4}{11}\right)^{4/3}\right]^{-1}\left( \frac{2+\frac{7}{8}\times 2\times 2}{61.75}\right)^{4/3} \approx 0.18.
    \label{eq:DeltaNeff_1}
\end{equation}

\noindent The $\Delta N_{\rm eff}$ value above has a small tension with the most recent value of $N_{\rm eff.}$ determined by Planck \cite{Planck:2018vyg}
\begin{equation}
     N_{\rm eff.}= 2.99_{-0.33}^{+0.34}~.
    \label{eq:PlanckNeff}
\end{equation}
A recent determination from the DESI collaboration including both CMB and BAO results in ~\cite{DESI:2024mwx}:
\begin{equation}
     N_{\rm eff.}=3.10 \pm 0.17~.
    \label{eq:DESINeff}
\end{equation}
Then, we see that the model is compatible with the $N_{\rm eff}$ data within $1~\sigma$. Future CMB experiments will have a much greater sensitivity to this quantity and therefore will be a significant test to this model.  
So, we conclude that having large twin neutrino masses and a decoupling temperature between the two QCDs is still compatible with the available data. One last comment concerns the role of the twin Majoron in $\Delta N_{\rm eff}$. For the benchmark point in Table~\ref{tab:pars}, the twin Majoron is already non-relativistic at the twin-SM decoupling. However, it is possible to have lighter twin Majorons by choosing a smaller scale for $b$. In Section~\ref{sec:DomainWalls} we comment on this possibility. 

\subsection{Collider Bounds}
\label{sec:LHCpheno}

We divide the collider phenomenology of the model into the usual searches in type-II seesaw models for EW triplets \cite{Ashanujjaman:2021txz,ATLAS:2017xqs,Aoki:2011pz,FileviezPerez:2008jbu,Fuks:2019clu} on the one hand, and the collider phenomenology of the twin sector \cite{Burdman:2014zta,Craig:2015pha} on the other. For the EW triplets, we look for direct production and decay of the extra scalars at the LHC and their indirect effects in the SM precision measurements. For the invisible twin states, collider experiments can probe modifications in the Higgs couplings induced by the heavy sector and put bounds on the invisible decay width of the Higgs to twin states. We summarize the constraints at the end of the section and describe the bounds in the following paragraphs. 

Direct searches for type-II seesaw triplets with EW quantum numbers are usually performed assuming the Drell-Yan production modes:
\begin{subequations}
    \begin{alignat}{5}
    & q \overline{q}'&& \rightarrow \hspace{2mm}{W^\pm}^* &&\rightarrow \Delta^{\pm\pm} \Delta^{\mp},&& \hspace{2mm} \Delta^{\pm} \Delta^0, && \hspace{2mm} \Delta_s^{\pm} \Delta_a^0, \\
    & q \overline{q}'&& \rightarrow \gamma^{*} / Z^{*} && \rightarrow \Delta^{\pm\pm} \Delta^{\mp\mp},&& \hspace{2mm} \Delta^{\pm} \Delta^{\mp}, && \hspace{2mm} \Delta_s^{0} \Delta_a^0,
\end{alignat}
\label{eq:prod}
\end{subequations}
\noindent where we dropped the $A$ labels of the fields as these are all SM states. Additionally, for non-zero $b$ coupling, the final states of \eqref{eq:prod} have a small doublet component due to mixing, which is relevant to describe their decays. From \eqref{eq:masses_tripletsA}, the coupling $\lambda_5$ controls the mass splitting of the visible triplets. However, $\lambda_5$ is generically small in our model since $\lambda_5 < \lambda'$ needs to be imposed to keep $F>f$. As discussed, the mixing of visible and twin states must be tuned as in \eqref{eq:tuning} to avoid a large invisible Higgs decay width. We comment on the mixing bounds later in this section, but the constraint \eqref{eq:tuning} leads to 
\begin{equation}
    \lambda_5 < \lambda' \ll \frac{\lambda v^2}{f^2}
\end{equation}
which implies 
\begin{equation}
    \Delta m^2 \equiv m_{\Delta_A^\pm}^2 - m_{\Delta_A^{\pm\pm}}^2 = m_{\Delta_A^0}^2 - m_{\Delta_A^{\pm}}^2 =\frac{\lambda_5}{4}v^2 \ll m_{\Delta_A^\pm}^2 .
\end{equation}

\noindent Therefore, the direct triplet searches relevant to us are the ones focused on the degenerate case, where $\Delta m^2 = 0$ is imposed. Generally, in this case, the decays of the scalars can be divided for a small triplet VEV scale, with $V \lesssim 10^{-4} \GeV$ and a large VEV, $V \gtrsim 10^{-4} \GeV$. The specific value of the triplet VEV, $10^{-4} \GeV$, divides parameter space, with the triplets decaying into leptonic final states for smaller $V$ and bosonic final states for larger $V$. This is due to the fact that the coupling of one triplet to two gauge bosons is proportional to the triplet VEV $V$, whereas their couplings to leptons just depend on $\kappa$ and $v$\footnote{For more details see Appendix A in Ref.~\cite{Ashanujjaman:2021txz}}. 
\begin{alignat}{5}
    & V \lesssim 10^{-4} \GeV: \Delta^{\pm\pm} \rightarrow l^\pm l^\pm, && \quad \Delta^{\pm}\rightarrow l^{\pm}\nu, \quad &&\Delta_{s,a} \rightarrow \nu\nu,  && \\
    & V \gtrsim 10^{-4} \GeV: \Delta^{\pm\pm} \rightarrow W^\pm W^\pm, && \quad \Delta^{\pm}\rightarrow W^{\pm}(Z/ h), \quad && \Delta_{s} \rightarrow \mathrm{VV}/ hh, && \quad \Delta_{a} \rightarrow hZ
\end{alignat}
where $\mathrm{V}=W^{\pm},Z$. The most constraining decay channels involve the doubly charged scalars $\Delta^{\pm\pm}$ as these produce multiple charged final states and are a distinct signal for the model. The analysis is divided between the multilepton searches for smaller triplet VEV and multiboson searches for larger $V$. We refer  to \cite{Ashanujjaman:2021txz} for a detailed description of the production and decay modes in the context of the type-II seesaw model, to the multilepton CMS analysis \cite{CMS:2019lwf} and to the multiboson ATLAS analysis \cite{ATLAS:2021jol}.

Using the $139 \text{ fb}^{-1}$ LHC dataset, in the case of $V \lesssim 10^{-4} \GeV$ and $\Delta m^2=0$, the CMS multiplepton analysis constrains the mass of the doubly charged scalar to be $m_{\Delta_A^{\pm\pm}} > 950 \GeV$ at 95\% confidence level. For $ V \gtrsim 10^{-4} \GeV$, the ATLAS multiboson analysis is able to constrain $m_{\Delta_A^{\pm\pm}} > 400 \GeV$ at 95\% confidence level.\footnote{The CMS and ATLAS bounds from Refs. \cite{CMS:2019lwf} and \cite{ATLAS:2021jol} used here correspond to the recasting performed for the type-II seesaw model in Ref.~\cite{Ashanujjaman:2021txz} using these data.} In both cases, our model can reproduce the allowed masses as indicated by the benchmark point of Table~\ref{tab:pars}.

The triplet scalars introduce a new source of custodial symmetry breaking at tree- and loop-level. This breaking manifests itself as a contribution to the W-boson mass and a modification of the SM $\rho$ parameter that enters several EW precision observables. For a triplet $(T,Y)=(1,2)$ and a doublet $(T,Y)=(1/2,1)$, the tree-level $\rho$ parameter is \cite{Gunion:1989we}
\begin{equation}
    \rho_{\rm tree} = \left.\frac{m_W^2}{m_Z^2 \cos\theta_W^2} \right|_{\rm tree} = \frac{{\sum}_{i} \left[ 4T_i(T_i+1)-Y_i^2\right]v_i^2}{{\sum}_{i}2Y_i^2 v_i^2}=1-\frac{2V^2}{v^2+4V^2}.
    \label{eq:rho_tree}
\end{equation}

\noindent The $\rho$ parameter is tightly constrained by LEP data to be $\rho=1.00038(20)$ \cite{ParticleDataGroup:2022pth,ALEPH:2013dgf}. Using \eqref{eq:rho_tree} and the experimental limit, we can bound the triplet VEV to be below $V\lesssim 1.1 \GeV$ at $95\%$ confidence level, which is orders of magnitude above what we need it to be to obtain the correct SM neutrino masses.

Additionally, $m_W^2$ gets a radiative contribution from the $\Delta^{\pm\pm}, \Delta^\pm$ intermediate state loops. Generically, the oblique contributions to the W-boson mass can be written as in \cite{Maksymyk:1993zm},
\begin{equation}
    m_W^2= m_{W {\rm ,SM}}^2 \left[1-\frac{\alpha_{\rm EM}S}{2(c_W^2-s_W^2)} + \frac{c_W^2 \alpha_{\rm EM}T}{(c_W^2-s_W^2)} + \frac{\alpha_{\rm EM}U}{4s_W^2} \right],
\end{equation}
where $S,T$ and $U$ are the Peskin-Takeuchi oblique parameters \cite{Peskin:1991sw} and $s_W^2=\sin^2\theta_W \simeq 0.23$. Including the triplet contributions, the oblique parameters are \cite{Mandal:2022zmy,Heeck:2022fvl}
\begin{align}
S_{\rm loop} &\simeq - \frac{(2-4 s_W^2+5 s_W^4) m_Z^2}{30\pi m_{\Delta^\pm}^2}+\lambda_5 \frac{v^2}{6\pi m_{\Delta^\pm}^2} \simeq  3\times 10^{-3}\frac{\lambda_5 - 0.04}{(m_{\Delta^\pm}/\TeV)^2} \,,
\label{eq:S}
\\
T_{\rm loop} &\simeq  \frac{ v^2 \lambda_5^2}{192\pi^2 \alpha_\text{EM}  m_{\Delta^\pm}^2} \simeq \frac{4\times 10^{-3}}{(m_{\Delta^\pm}/\TeV)^2}\,\lambda_5^2\,, 
\label{eq:T}
\\
U_{\rm loop} &\simeq  \frac{(2-4 s_W^2 +5 s_W^4) m_Z^2 - 2 m_W^2}{30\pi m_{\Delta^\pm}^2} \simeq \frac{-2\times 10^{-5}}{(m_{\Delta^\pm}/\TeV)^2}\,.
\label{eq:U}
\end{align}

\noindent As it is usual, the $U$ parameter contribution is more suppressed than the other two since it is associated to a dimension eight operator. The global fit reported by the PDG \cite{ParticleDataGroup:2022pth} constrains the values of the oblique parameters to be $S=-0.02\pm 0.10$, $T=0.03\pm 0.12$ and $U=0.01\pm0.11$\footnote{Recent measurements by the CDF collaboration suggest that the best-fit value is $S=(0.17, 0.27)$ assuming $U=0$, which is in tension with the previous analysis. This result was interpreted in the context of type-II seesaw models \cite{Heeck:2022fvl,Ghosh:2022zqs,Butterworth:2022dkt}. However, for our model, the corrections have the opposite sign to explain the CDF best-fit value.}.We can see from (\ref{eq:S}), (\ref{eq:T}) and (\ref{eq:U}) that the loop contributions in this model to $S$, $T$ and $U$ are negligibly small, provided that the triplet scalars live close to the $\TeV$ scale.

Going to the twin sector, collider bounds are similar to the ones imposed in the usual implementations of the twin Higgs mechanism \cite{Batell:2022pzc}. The vacuum misalignment between the two doublet VEVs induces a modification of the Higgs couplings to fermions and gauge bosons. The present experimental limits, derived from the Higgs coupling measurements at the LHC \cite{ATLAS:2021vrm,CMS:2022dwd}, constrain $f/v$ to be greater than approximately 3 \cite{Burdman:2014zta}. Additionally, measurements on the invisible Higgs decay width, \cite{ATLAS:2022yvh,CMS:2022qva}, constrain the mixing with the twin states via the Higgs portal interactions. The invisible Higgs width is constrained by \cite{ATLAS:2022yvh} to be smaller than $18\%$, and by \cite{CMS:2022qva} to be smaller than $14\%$. Using the latter value, requires $f \gtrsim 2.7 v$, which can be translated to a tuning in the parameters of the off-diagonal terms of the CP-even Higgs mass matrix \eqref{eq:M2_CPeven}. Because of this, we need to impose the conditions
\begin{align}
    2 \lambda +\lambda' \lesssim \frac{v^2}{f^2} \sim 0.1, \hspace{1cm} \lambda_4 \eta \lesssim \frac{v^2}{f^2} \sim 0.1.
\end{align}

Finally, we summarize the bounds from accelerator experiments,
\begin{alignat}{5}
    &\mathbf{m_{\Delta_A^{\pm\pm}} \geq 950 \textbf{ GeV}} && \text{ at $95\%$ C.L. for } V \lesssim 10^{-4} \GeV \text{ (CMS \cite{CMS:2019lwf})}, 
    \\
    &\mathbf{m_{\Delta_A^{\pm\pm}} \geq 400 \textbf{ GeV}} && \text{ at $95\%$ C.L. for } V \gtrsim 10^{-4} \GeV \text{ (ATLAS \cite{ATLAS:2021jol})},
\end{alignat}
\begin{equation}
    \text{(LFVT \cite{Cheng:2022jyi})   } \mathbf{ 0.7 \mathbf{eV} \leq V \leq 1.1 \mathbf{GeV}} \text{    (EWPT \cite{Ashanujjaman:2021txz})} \text{ at $95\%$ C.L.}.
\end{equation}
LFVT are the lepton flavor violation tests discussed in the previous section, and EWPT are the electroweak precision tests discussed above.

\subsection{Twin Majoron Bounds and Domain Wall Decays}
\label{sec:DomainWalls}

We now turn to two other aspects of the cosmological history of the model. First, we derive bounds on a light relativistic twin Majoron, and then we address the potential problem with the formation of a domain wall network due to the spontaneous breaking of a discrete symmetry, here the $\Z2$ exchange symmetry. 

The phenomenology of a SM triplet Majoron has been extensively studied in the literature~\cite{Ma:1998dx,Ma:2017xxj,Gelmini:1980re,Schechter:1981cv,Ma:2000xh,Diaz:1998zg,deSPires:2003wwk}, and results in constrains mostly from measurements of the $Z$ boson width. However, in our case, the Majoron is only part of the twin sector where there are no bounds from EW precision measurements. Therefore, most of the constraints on the twin Majoron arise from its thermal history. In Section~\ref{sec:neutrinomasses}, we computed $\Delta N_{\rm eff}$ without including the twin Majoron. This was justified for the benchmark value in Table~\ref{tab:pars}, since $J_B$ is non-relativistic by the time of SM-twin sector decoupling. However, for lower values of $b$ and higher $\kappa$, the twin Majoron becomes lighter and can be relativistic at decoupling or BBN and CMB times. Thus, we need to reconsider the case of a  lighter twin Majoron, and compute its lifetime and decay channels to estimate its effect on cosmology. 

We can integrate out the heavy degrees of freedom above the $\GeV$ scale to obtain the effective Lagrangian for the Majoron
\begin{equation}
    \mathcal{L}_{J_B}= c_\gamma J_B F_{\mu\nu}\widetilde{F}^{\mu\nu} + c_{\psi,i} J_B \overline{\psi}_i \gamma_5 \psi_i + c_{\nu,ij} J_B \overline{\nu_i^c} \nu_j,
\end{equation}
where $\psi = (\psi_L, \psi_R)^T$ are Dirac spinos corresponding to the electron, muon, and first generation quarks. The couplings of the Majoron with the photon and gauge bosons are generated from the twin Higgs Yukawa coupling due to the mixing with the pseudoscalar doublet $G_B^0$. In terms of the parameters of the potential, we can write 
\begin{align}
    &(c_\nu)_{ij} = \kappa_{ij} \cos\alpha = \frac{2F \kappa_{ij}}{\sqrt{f^2+4F^2}}\\
    &(c_\psi)_i = y_{\psi_i} \sin\alpha = \frac{f y_{\psi_i}}{\sqrt{f^2+4F^2}}, 
\end{align}
where $y_{\psi_i}$ is the $i^{\rm th}$ light fermion Yukawa coupling and $\alpha$ the mixing angle defined in \eqref{eq:majoron_alpha}. The photon coupling comes from the loops of the charged fermionic fields,
\begin{equation}
    c_\gamma = \frac{e^2}{32\pi^2} {\sum}_i \frac{y_{\psi_i}\sin\alpha}{m_{\psi_i}} = \frac{N_\psi e^2}{32\pi^2 v^2} \frac{f}{\sqrt{f^2+4F^2}}.
\end{equation}

\noindent Here, $N_\psi$ is the number of charged fermions that are heavier than the Majoron. Then, the decay rate of each channel is given by
\begin{align}
    &\Gamma(J_B\rightarrow \nu_B\nu_B)= \frac{\sum_{ij} |\kappa_{ij}|^2 m_{J_B}}{8\pi}  \frac{F^2}{f^2+4 F^2} \sqrt{1-\frac{4m_{\nu_B}^2}{m_{J_B}^2}}, \\
    &\Gamma(J_B\rightarrow l^+_{B,i} l^-_{B,i})=\frac{y_l^2 m_{J_B}}{4\pi}\frac{f^2}{f^2+4F^2}\sqrt{1-\frac{4m_{l_{B,i}}^2}{m_{J_B}^2}},\\
    &\Gamma(J_B\rightarrow \gamma_B\gamma_B) = \frac{N_\psi^2\alpha_e^2}{64\pi^3}\frac{m_{J_B}^3}{v^2}\frac{f^2}{f^2+4F^2}.
\end{align}

\noindent
Out of the possible Majoron decay channels, the smallest partial width is twin electrons and muons, as these are proportional to the lepton Yukawa couplings. The neutrino decay can be sizable depending on the value of $\kappa$. However, as discussed below, the twin neutrinos are heavier than the twin Majoron for most of the allowed parameter space, making the decay channel forbidden. Therefore, we conclude that the twin Majoron decays to photons with a branching ratio of approximately one. As a result, the Majoron lifetime can be approximated by 
\begin{equation}
    \tau_{J_B} \simeq 1 \text{s} \left(\frac{1.1\MeV}{m_{J_B}}\right)^3\left(\frac{\xi}{0.1}\right)\frac{f^2+4F^2}{(2.1\TeV)^2}.
\end{equation}

\noindent Thus, a Majoron that is lighter than $1 \MeV$ would decay during BBN, injecting entropy into the twin sector. This would lead to an increase in $\Delta N_{\rm eff}$ from the entropy injections, which would effectively increase the twin sector temperature. In our case, we consider $m_{J_B} \gtrsim 1 \MeV$, so the contributions due to entropy injection are unimportant. However, we must consider the contribution to $\Delta N_{\rm eff}$ from having relativistic twin Majorons and neutrinos during and after the SM-twin decoupling.

Previously, we assumed that twin Majoron and twin neutrinos were non-relativistic by the time of decoupling between the SM and twin sectors. This rendered $\Delta N_{\rm eff}\simeq 0.18$ as shown in \eqref{eq:DeltaNeff_1}. But if the Majoron is much lighter than the twin decoupling temperature but heavier than $\sim 1 \MeV$, it contributes to $g_{*,S}^B(T_{dec})$, increasing the temperature ratio that enters \eqref{eq:DeltaNeff_def}. Therefore, we have
\begin{equation}
    \Delta N_{\rm eff} \simeq 4.4\left( \frac{2+\frac{7}{8}\times 2\times 2+1}{61.75}\right)^{4/3} \approx 0.21,  \quad (1\MeV \lesssim m_{J_B} \lesssim 100 \MeV).
    \label{eq:DeltaNeff_2}
\end{equation}

\begin{figure}
    \centering
    \includegraphics[scale=0.26]{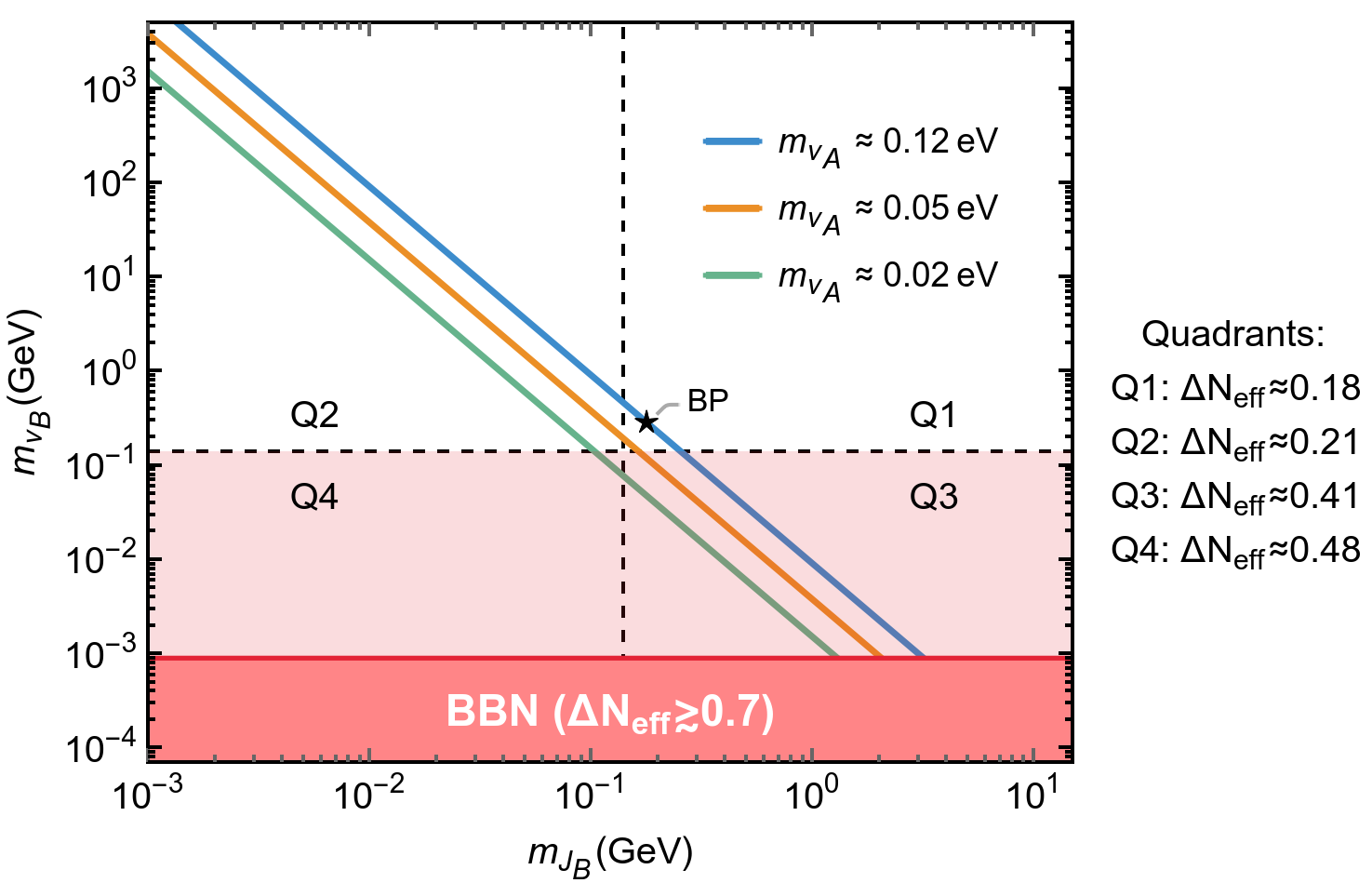}
    \caption{Twin Majoron against twin neutrino masses for fixed values of SM neutrino masses. The chosen benchmark point (BP) from Table \ref{tab:pars} gives the most favorable scenario for $\Delta N_{\rm eff} \simeq 0.18$. Each quadrant reflects calculated $\Delta N_{\rm eff}$ values assuming relativistic/non-relativistic twin Majoron and twin neutrinos. The exclusion bound from BBN comes from late twin neutrino decays that increase $\Delta N_{\rm eff}$ to values greater than $0.7$, which is incompatible with data \cite{ParticleDataGroup:2022pth,DESI:2024mwx}.}
    \label{fig:mJB_mnuB}
\end{figure}

\noindent If instead of the Majoron, the neutrinos are relativistic at decoupling, $\Delta N_{\rm eff} \simeq 0.41$. Alternatively, if both particles are relativistic below the decoupling temperature, then $\Delta N_{\rm eff} \simeq 0.48$. Finally, suppose the twin neutrinos are lighter than roughly the BBN temperature. In that case, their decays will reheat the twin sector rendering a large effective number of extra neutrino species, $\Delta N_{\rm eff} \simeq 0.7$. While having a light twin Majoron with $\Delta N_{\rm eff} \simeq 0.21$ is still not excluded by data~\cite{Planck:2018vyg,DESI:2024mwx}, having twin neutrinos below the decoupling temperature is in severe tension with cosmological data. In Figure~\ref{fig:mJB_mnuB}, we plot the twin Majoron and twin neutrino masses for fixed values of the SM neutrino masses. The benchmark point chosen in Table~\ref{tab:pars} gives the best case scenario for $\Delta N_{\rm eff} \simeq 0.18$, as both light twin particles are non-relativistic. Each quadrant corresponds to the $\Delta N_{\rm eff}$ values calculated above. 

Finally, we address the issue of the formation of domain walls in the model. Because of the model's exact $\Z2$ symmetry, the other vacua obtained by changing the values of $v \rightarrow f$ and $V \rightarrow F$ have the same depth as the one we considered. Therefore, as the universe cools down, the scalar fields can relax to any of these equivalent vacua, potentially leading to a domain wall problem \cite{Zeldovich:1974uw}. The formation of the domain wall network is model-dependent as we need to specify the inflationary and reheating dynamics during the radiation domination era. As assumed in \cite{Beauchesne:2021opx}, if the Hubble parameter during inflation and the reheating temperature are below the model's phase transition scale, then the domain walls are not formed, and the model is cosmologically consistent. On the other hand, if we suppose inflation and reheating happened before the twin-phase transition, we need a mechanism to collapse the different domains to the vacuum considered throughout the paper, i.e. \eqref{eq:vevs}. 

As shown in Ref.~\cite{Batell:2019ptb}, one way to achieve the collapse of the domain wall network is to introduce a very small amount of explicit $\Z2$ breaking. For example, we can assume the following soft-$\Z2$ breaking term
\begin{equation}
    V_{\cancel{\Z2}}= \mu_1^2 \big( |H_A|^2 - |H_B|^2 \big) + \mu_2^2 \big( \tr\Delta_A^{\dagger}\Delta_A - \tr\Delta_B^{\dagger}\Delta_B \big),
\end{equation}

\noindent where $\mu_{1,2}^2 \ll |m^2|, |M^2|$ is a positive $\Z2$ constant. This term effectively selects the vacuum we are interested in by lowering the value of the potential at the correct minimum. The potential contribution in the true minimum, which is defined in \eqref{eq:vevs}, is 
\begin{equation}
    \left.\langle V_{\cancel{\Z2}} \rangle \right|_{\rm true}= \mu_1^2 \big( v^2 - f^2 \big) + \mu_2^2 \big( V^2 - F^2 \big) < 0,
\end{equation}

\noindent We can obtain the $\Z2$ symmetric minimum by changing $v\leftrightarrow f$ and $V\leftrightarrow F$. The contribution to the potential is
 \begin{equation}
    \left.\langle V_{\cancel{\Z2}} \rangle \right|_{\rm false}= \mu_1^2 \big( f^2 - v^2 \big) + \mu_2^2 \big( F^2 - V^2 \big) > 0.
\end{equation}

\noindent Therefore, the true minimum has the correct VEV hierarchy, $V\ll v< f \lesssim F$ needed for our model. Following \cite{Vilenkin:1981zs,Batell:2019ptb}, we can estimate how much the potential has to be shifted to make the false domains decay before they can dominate the energy density of the early universe. This requires 
\begin{equation}
    \Delta V \gtrsim \frac{\sigma^2}{M_{\rm pl}^2} 
    \label{eq:DeltaV}
\end{equation}

\noindent where $\sigma$ is the surface energy density of the wall, and $M_{\rm pl} \simeq 1.2 \times 10^{19} \GeV$. The surface density can be estimated as $\sigma \sim V_0^{1/2}f^2 $, where $V_0\sim \lambda f^2$ or $V_0\sim \lambda_1 F^2$ is the height of the potential barrier between the two minima \cite{Kitajima:2015nla}, depending on which quartic term is larger. Then, we can extract a bound on the $\Z2$ breaking parameters $\mu_{1,2}$ by imposing \eqref{eq:DeltaV}, with the shift in the potential given by $\Delta V \simeq 2 \left.\langle V_{\cancel{\Z2}} \rangle \right|_{\rm true}$.
\begin{equation}
    \mu_{1,2}^2\gtrsim \max \left[\frac{\lambda f^4}{M_{\rm pl}^2},\frac{\lambda_1 F^4}{M_{\rm pl}^2} \right].
\end{equation}

\noindent Imposing the condition (\ref{eq:DeltaV}), we finally obtain that $\mu_{1,2}\simeq 0.1 ~\text{meV}$ or bigger. Thus, we see that the explicit $\Z2$ breaking needed to be introduced is orders of magnitude smaller than any scale in the problem and it would have no observable consequences in either the spectrum or the couplings of the model presented here. Therefore, we see that it is justified to completely ignore the explicit $\Z2$ breaking effects, and the way the model was presented is enough to make sensible predictions without the need to consider effects of order $\mu_{1,2}/\Lambda$, where here $\Lambda$ stands for any other scale in the theory, i.e. $v$,$f$, $F$, etc.

\section{Conclusions}
\label{sec:conc}

We presented an extension of the mirror twin Higgs model where the $\Z2$ symmetry that relates the SM particles to the mirror sector is spontaneously broken. To achieve $\Z2$ spontaneous symmetry breaking, we introduce two $SU(2)$ triplets, one for the SM and one for the twin sector. The parameters of the scalar potential are fixed to give a particular hierarchy of VEVs, with $V\ll v < f \lesssim F$, where $V=\langle \Delta_A \rangle, v=\langle H_A \rangle, f=\langle h_B \rangle$ and $F=\langle \Delta_B \rangle$. The $SU(2)_{A,B}$ triplets are introduced to implement the type-II seesaw mechanism to give SM and twin neutrino masses. The masses of the neutrinos are proportional to the triplet VEV of the corresponding sector. Because of the VEV hierarchy obtained, the SM neutrinos acquire small masses proportional to $V$, while the twin neutrinos are much heavier, at the scale $F$. The lifting of the twin neutrino masses considerably ameliorates potential conflict with the cosmological bounds on extra radiation during BBN and CMB times. If twin neutrinos are heavier than $\sim 100 \MeV$ they do not contribute to the number of relativistic degrees of freedom at decoupling of the SM and twin sectors. The resulting contribution to the effective number of extra neutrino species is $\Delta N_{\rm eff}\simeq 0.18$, given almost exclusively by the twin photon contribution. This is within $1\sigma$ confidence level of the most recent measurements \cite{Planck:2018vyg,ParticleDataGroup:2022pth,DESI:2024mwx}.

One implication of the proposed mechanism is the appearance of a twin Majoron, the pNGB of spontaneously broken twin lepton number. The Majoron does not appear in the SM sector since the explicit breaking of SM lepton number is of the same order as the triplet VEV, $V$. For the twin sector, the explicit breaking induced by the triplet scalar coupling $b$ is much smaller than $F$. The most relevant constraints on the twin Majoron come from bounds on $\Delta N_{\rm eff}$, as discussed in Section~\ref{sec:DomainWalls}. For a twin Majoron lighter than $\sim 100 \MeV$, there is an additional contribution to the effective number of relativistic degrees of freedom at decoupling, which renders $\Delta N_{\rm eff} \simeq 0.21$. While slightly larger than the best-case scenario with non-relativistic twin Majorons, $\Delta N_{\rm eff}$ is still within $1\sigma$ confidence level from the measurements cited above.

As emphasized in Section~\ref{sec:pheno}, the model has several distinct experimental signatures, ranging from neutrino and collider phenomenology to cosmology. Neutrino oscillation data and neutrino mass measurements constrain the SM sector's triplet couplings to neutrinos and triplet VEV scale. Additionally, SM neutrino masses determine the relationship between the twin Majoron and twin neutrinos, as indicated in Figure~\ref{fig:mJB_mnuB}. Then, we can calculate the additional contributions to $\Delta N_{\rm eff}$ as discussed in the previous section. Regarding collider phenomenology, the main constraints come from direct searches for visible $SU(2)$ triplets and bounds on the Higgs couplings and mixing. Direct searches for the doubly charged Higgs constrain $\Delta_A^{\pm\pm}$ mass to be larger than $950 \GeV$ for $V\lesssim 10^{-4}\GeV$, and larger than $400 \GeV$ for $V \gtrsim 10^{-4}\GeV$. Bounds from the $\rho$ parameter restrict the triplet VEV to be $V \lesssim 1 \GeV$, which is orders of magnitude above what is necessary for realistic SM neutrino masses. On the other hand, lepton flavor violation tests require a lower bound of $V \gtrsim 0.7 \eV$ to best fit the oscillation parameter data. Additionally, as usual in twin Higgs models, the combination of the doublet VEVs must satisfy $f/v\gtrsim 3$ to be compatible with measurements of the Higgs couplings. Finally, in cosmology, if the scale of inflation is much above the scale of EW and $\Z2$ spontaneous symmetry breaking, then a small source of $\Z2$ explicit breaking must be introduced to avoid a domain wall problem. We have shown that the necessary explicit $\Z2$ breaking parameter is much smaller than any other scale in the problem and, therefore, does not change any of the predictions of the model in any observable way. 

In sum, the model has several correlated signals from various sources, including neutrino oscillation data, collider searches, and cosmological bounds. The future of these experiments~\cite{CMB-S4:2016ple,DUNE:2022aul,Hyper-Kamiokande:2018ofw,Mlynarikova:2023bvx} could significantly test this particular implementation of the twin Higgs model. From the theoretical side, we have shown that the mirror twin Higgs is very versatile, adeptly accommodating both the spontaneous breaking of the $\Z2$ symmetry and the generation of neutrino masses. It remains one of the most compelling models of physics beyond the standard model that addresses the hierarchy problem. 

\acknowledgments
The authors thank Kfir Blum for early collaboration in this project. We also acknowledge the support of FAPESP grant 2019/04837-9 and CAPES grant 88887.816450/2023-00. 

\bibliography{twinNu}
\bibliographystyle{JHEP}

\end{document}